# Experimental and theoretical research of photoneutron reactions in the $^{181}$Ta nucleus

J.H. Khushvaktov [a,b], M.A. Demichev [a], D.L. Demin [a], S.A. Evseev [a], M.I. Gostkin [a], V.V. Kobets [a], D.V. Ponomarev [a], F.A. Rasulova [b,], S.V. Rozov [a], E.T. Ruziev [b], A.A. Solnyshkin [a], T.N. Tran [a,c,d], E.A. Yakushev [a]

[a] *Joint Institute for Nuclear Research (JINR), Dubna, Russia*

[b] *Institute of Nuclear Physics of the Academy of Sciences of the Republic of Uzbekistan (INP ASRU), Tashkent, Uzbekistan*

[c] *Institute of Physics of the Vietnam Academy of Science and Technology (IP VAST), Hanoi, Vietnam*

[d] *Institute for Nuclear Science and Technology, VINATOM, 179 Hoang Quoc Viet, Cau Giay, Hanoi, Vietnam*

**Abstract**

Bremsstrahlung fluxes for irradiating tantalum samples were formed by irradiating a tungsten converter with an electron beam with energy up to 130 MeV. The relative yields and flux-averaged cross-sections of multinucleon photonuclear reactions with the emission of up to 9 neutrons in $^{181}$Ta nuclei were determined. Monte Carlo simulations to study the yields of photonuclear reactions were performed using Geant4 and TALYS-2.0 codes. The obtained experimental results were compared with the available literature data and calculated results. The comparison showed that the values of the relative reaction yield and the flux-averaged cross-section coincide with the literature data, taking into account the different geometry of the experiments. The calculated results coincide with the experimental ones only for reactions with the emission of up to 5 neutrons from the nucleus.

**Keywords:** photonuclear reaction, cross section, yield, neutron, simulation, Geant4, TALYS

## 1. Introduction

In most studies of photonuclear reactions, using bremsstrahlung, the region of giant dipole resonance (GDR) was studied, in which the nucleus has a highly excited state with the participation of a large number of nucleons. At the same time, no less interesting is the region lying behind the GDR maximum and extending up to the meson threshold (135 MeV), in which the photon predominantly interacts with quasi-deuterons formed inside the nucleus, and ends with the emission of several (up to ten) nucleons by the nucleus. This is due to the change in the mechanism of interaction of photons with nuclei from the excitation of GDR to the quasi-deuteron mechanism [1,2]. To describe the mechanism of multi-particle photonuclear reactions, various theoretical models have been developed and tested by comparison with experimental literature data. For example, a combined model of photonucleon reactions, which combines a semi-empirical model of oscillations, a quasi-deuteron model of photoabsorption, an exciton and evaporation model [3-5]. And also the Monte Carlo model of a multi-collision intranuclear cascade which can describe photonuclear reactions at intermediate energies from 20 to 140 MeV [6,7]. Despite the above studies, photonuclear reactions in tantalum nuclei in the energy range above 30 MeV have been little studied. To verify the correctness of theoretical models of the quasi-deuteron mechanism, the database on the yields and cross-sections of photonuclear reactions in most stable nuclei, including tantalum nuclei, needs to be supplemented with new experimental data. In this work, the processes of interaction of bremsstrahlung with $^{181}$Ta nuclei in the range of end-point energies from 20 to 130 MeV are studied experimentally and theoretically.

## 2. Experimental design and data analysis

The experiments have been carried out at the LINAC-200 electron accelerator [8]. A tungsten converter with a size of 4.5*4.5*0.5 cm was irradiated with electron beams with energies of 20, 40, 60, 80, 105, and 130 MeV. The diameter of the electron beam incident on the converter is 5.5±0.5 mm. Behind the tungsten converter were tantalum samples with masses of 142, 135, 163, 718, and 207 mg, respectively, for experiments with an electron beam with energies of 20, 60, 80, 105, and 130 MeV. Tantalum samples were irradiated with a bremsstrahlung flux generated in a tungsten converter in the experiments described above, except for the experiment with electrons with an energy of 40 MeV. And in the experiment with an electron beam with an energy of 40 MeV, a set of five samples (Se, Co, Y, Tb and Ta) were irradiated with a direct electron beam. The mass of the tantalum sample was 184 mg; it was the last one in the set of samples and was irradiated with the bremsstrahlung flux generated in the Se, Co, Y, and Tb samples. The irradiation time was 40, 20, 25, 15, 20, and 15.5 min, and the mean current was 0.40, 0.96, 0.80, 0.80, 1.16, and 1.00 μA, respectively, with electron energy. The current was measured using an inductive current sensor based on a Rogowski coil, the measurement uncertainty using the sensor is less than 2%.

After irradiation, the tantalum samples were transferred to the test room, and their gamma spectra were measured using an HPGe detector (model GR1819). More than ten gamma spectra of each sample were measured with different measurement times. The time from the end of irradiation to the start of measuring the first spectrum of the sample was 163, 27, 21, 23, 20 and 34 min, respectively, with electron energy. The gamma spectra obtained were processed using the Deimos32 program [9]. The areas of the identified peaks were determined while the background from the Compton scattering of photons was subtracted. Figure 1 shows the gamma spectrum of a tantalum sample from an experiment with 130 MeV electrons and the background spectrum at the measuring location. The spectrum of the tantalum sample was measured for 1 hour, and the time after irradiation until the start of measuring this spectrum was 4 hours. The background spectrum is also normalized to 1 hour measurement time. The figure 1 also shows examples of gamma peak identification. The absolute efficiency of the HPGe detector was measured using standard gamma sources at the same distances from the detector at which the tantalum samples were examined. Figure 2 shows the results of measuring the absolute efficiency for a distance from the detector of 1.3-13.5 cm. When processing the experimental data, we used the interpolation function of the measured values of the detector efficiency.

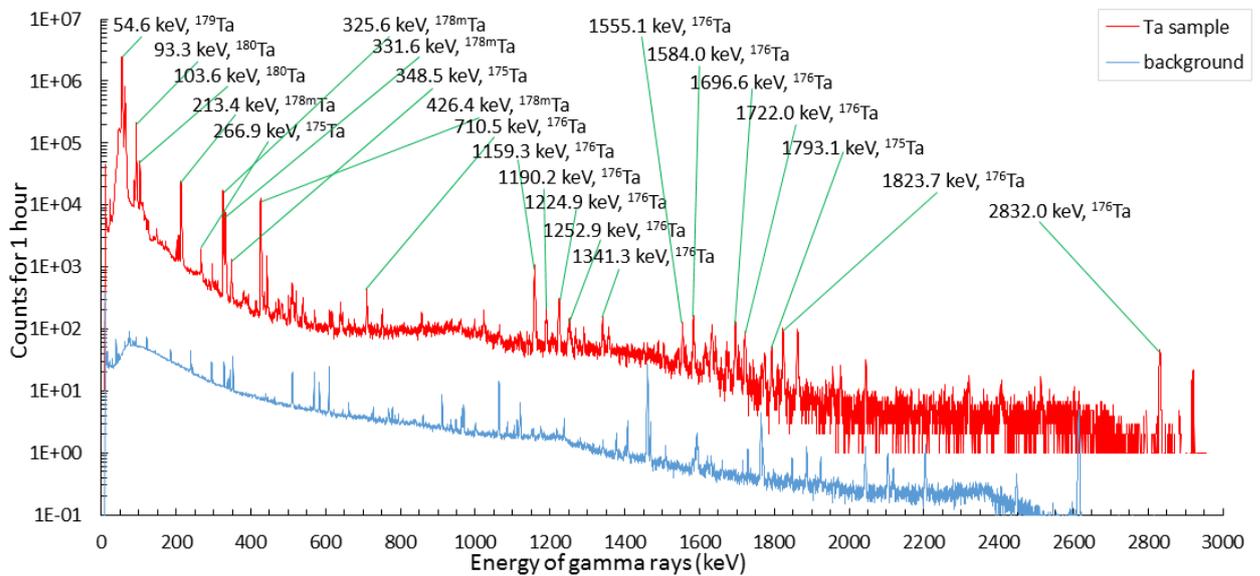

Fig.1. Gamma spectrum of a tantalum sample and background spectrum at the measurement location.

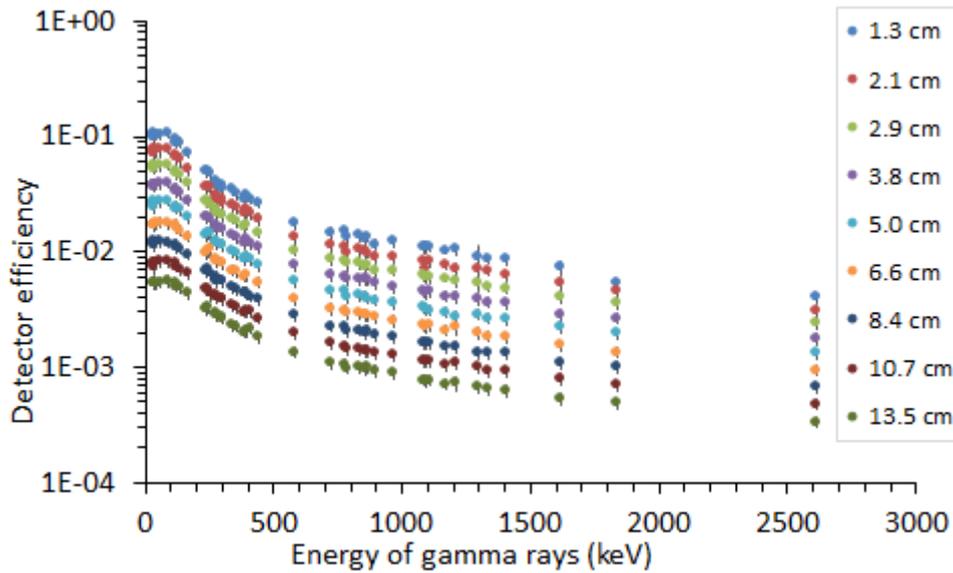

Fig.2. The absolute efficiency of the HPGe detector.

The yields of photonuclear reactions in the samples can be determined with the following formula:

$$Y_{exp} = \frac{S_p \cdot C_{abs}}{\varepsilon_p \cdot I_\gamma} \frac{t_{real}}{t_{live}} \frac{1}{N} \frac{1}{N_\gamma} \frac{e^{\lambda \cdot t_{cool}}}{1-e^{-\lambda \cdot t_{real}}} \frac{\lambda \cdot t_{irr}}{1-e^{-\lambda \cdot t_{irr}}}, \quad (1)$$

where $S_p$ is the full-energy peak area, $\varepsilon_p$ is the full-energy peak detector efficiency, $C_{abs}$ is the self-absorption correction coefficient, $I_\gamma$ is the gamma emission probability, $t_{real}$ and $t_{live}$ are the real time and the live time of the measurement, respectively, $N$ is the number of atoms in a sample, $N_\gamma$ is the integral number of photons incident on the tantalum sample, $\lambda$ is the decay constant, $t_{cool}$ is the cooling time, and $t_{irr}$ is the irradiation time. But due to the impossibility of accurate measurements of the $N_\gamma$ total number of photons, we were able to determine the ratio of yields of photoneutron reactions. When determining the ratio of the reaction yields, the number of atoms in the sample and the total number of photons incident on the sample will be cancelled. In this work, the ratio of the yield of photoneutron reactions in $^{181}$Ta nuclei with the release of two or more neutrons to the yield of a reaction with the release of a single neutron is determined.

The values of the parameters of the nuclear reactions studied in this work according to data from the [10] are given in the Table 1. $E_{th}$ are reaction thresholds, J, π, and $T_{1/2}$ are the spin, parity, and half-life of the nuclear reaction products, respectively, $E_\gamma$ are the energies of gamma rays emitted by the reaction products, $I_\gamma$ are the intensity of gamma rays. The values of the reaction thresholds $E_{th}$ are obtained from the TALYS-2.0 program [11]. For all identified gamma rays of radionuclides listed in Table 1, the ratio of the reaction yields were calculated; if a radionuclide is identified with more than one gamma line, the values of the ratio of reaction yields for them are averaged.

Table 1. Spectroscopic data [10] on product nuclei of measured photonuclear reactions.

| Nuclear reaction | $E_{th}$ (MeV) | $J^\pi$ of nucleus-product | $T_{1/2}$ | $E_\gamma$ (keV) | $I_\gamma$ (%) |
|---|---|---|---|---|---|
| $^{181}$Ta($\gamma$,n)$^{180}$Ta | 7.64 | $1^+$ | 8.15 (1) h | 93.326 (2) | 4.51 (16) |
| | | | | 103.557 (7) | 0.87 (16) |
| $^{181}$Ta($\gamma$,2n)$^{179}$Ta | 16.73 | $7/2^+$ | 1.82 (3) y | 55.786 | 22.09 (19) |
| | | | | 54.608 | 12.62 (12) |
| | | | | 63.000 | 7.29 (13) |
| $^{181}$Ta($\gamma$,3n)$^{178m}$Ta | 22.36 | $7^-$ | 2.36 (8) h | 426.383 (6) | 97.0 (13) |
| | | | | 325.562 (4) | 94.1 (11) |
| | | | | 213.440 (3) | 81.4 (11) |
| | | | | 88.867 (1) | 64.4 (10) |
| | | | | 331.613 (9) | 31.19 (19) |
| $^{181}$Ta($\gamma$,3n)$^{178}$Ta | 22.15 | $(1)^+$ | 9.31 (3) min | 1350.68 (3) | 1.18 (3) |
| | | | | 1340.8 (2) | 1.027 (24) |
| $^{181}$Ta($\gamma$,4n)$^{177}$Ta | 29.03 | $7/2^+$ | 56.36 (13) h | 112.950 (1) | 7.2 (8) |
| $^{181}$Ta($\gamma$,5n)$^{176}$Ta | 37.70 | $(1)^-$ | 8.09 (5) h | 1159.28 (9) | 24.7 (18) |
| | | | | 201.83 (3) | 5.7 (4) |
| | | | | 1224.93 (7) | 5.7 (4) |
| | | | | 710.50 (8) | 5.4 (4) |
| | | | | 1584.02 (10) | 5.3 (4) |
| | | | | 1696.55 (13) | 4.6 (3) |
| | | | | 1190.22 (10) | 4.5 (3) |
| | | | | 1823.70 (15) | 4.5 (3) |
| | | | | 2832.0 (2) | 4.3 (3) |
| | | | | 1555.07 (10) | 4.0 (3) |
| | | | | 1341.33 (10) | 3.3 (2) |
| | | | | 1722.04 (13) | 3.27 (24) |
| | | | | 1252.87 (10) | 3.08 (23) |
| $^{181}$Ta($\gamma$,6n)$^{175}$Ta | 44.86 | $7/2^+$ | 10.5 (2) h | 348.5 (5) | 12.0 (6) |
| | | | | 266.9 (4) | 10.8 (13) |
| | | | | 1793.1 (3) | 4.6 (6) |
| | | | | 436.4 (7) | 3.8 (2) |
| | | | | 857.7 (3) | 3.2 (3) |
| | | | | 998.3 (4) | 2.6 (3) |
| | | | | 393.2 (6) | 2.12 (16) |
| | | | | 475.0 (7) | 2.04 (20) |
| $^{181}$Ta($\gamma$,7n)$^{174}$Ta | 53.39 | $3^+$ | 1.14 (8) h | 206.50 (4) | 60 (5) |
| $^{181}$Ta($\gamma$,8n)$^{173}$Ta | 61.14 | $5/2^-$ | 3.14 (13) h | 172.2 (1) | 17.5 (18) |
| $^{181}$Ta($\gamma$,9n)$^{172}$Ta | 70.00 | $(3)^+$ | 36.8 (3) min | 1109.27 (9) | 14.9 (15) |
| | | | | 1330.41 (6) | 8.1 (8) |

### 3. Monte Carlo simulations

In simulations using the Geant4 code, classes G4eBremsstrahlung, G4PenelopeBremsstrahlung and G4LivermoreBremsstrahlungModel calculate the energy loss of electrons and positrons due to the radiation of photons in the nuclear field. The classes G4eBremsstrahlung, G4PenelopeBremsstrahlung and

G4LivermoreBremsstrahlungModel are based on the Seltzer-Berger bremsstrahlung model, Penelope Model and Livermore Model, respectively. In the above models, below electron energies of 1 GeV, the cross section evaluation is based on a dedicated parameterization, above this limit an analytic cross section is used [12]. In our simulations we used the class G4eBremsstrahlung (default class).

The Seltzer-Berger bremsstrahlung model was developed based on interpolation of tables of differential cross sections [13,14], covering electron energies from 1 keV to 10 GeV. Single-differential cross section can be written as a sum of a contribution of bremsstrahlung produced in the field of the screened atomic nucleus $\frac{d\sigma_n}{dk}$, and the part $Z\frac{d\sigma_e}{dk}$ corresponding to bremsstrahlung produced in the field of the Z atomic electrons,

$$\frac{d\sigma}{dk} = \frac{d\sigma_n}{dk} + Z\frac{d\sigma_e}{dk}.$$

The differential cross section depends on the energy $k$ of the emitted photon, the kinetic energy of the incident electron and the atomic number $Z$ of the target atom.

Due to the difficulty of using a sufficient number of electrons to determine the number of photonuclear reactions with a small error, we were able to obtain only the bremsstrahlung fluence in calculations with Geant4. Further, the yields of the photonuclear reactions were determined using formula (2). And the reaction cross sections were calculated using the TALYS-2.0 program. Statistical models use nuclear level densities at excitation energies to predict cross sections when information about discrete levels is unavailable or incomplete. Several level density models can be used in TALYS, from phenomenological analytical expressions to tabulated level densities derived from microscopic models [11]. In cross section calculations we used ldmodel 1 (Constant Temperature + Fermi gas model, default model).

$$Y_{calc} = \int_{E_{thr}}^{E_{max}} f(E)\sigma(E)dE, \qquad (2)$$

where $f(E)$ is the bremsstrahlung fluence, $\sigma(E)$ is the reaction cross section. Figure 3 shows the bremsstrahlung fluence incident on a tantalum sample in experiments with electrons with energies of 20, 60, 80, 105 and 130 MeV. Bremsstrahlung with energies above 7.6 MeV are capable of causing photoneutron reactions in nuclei.

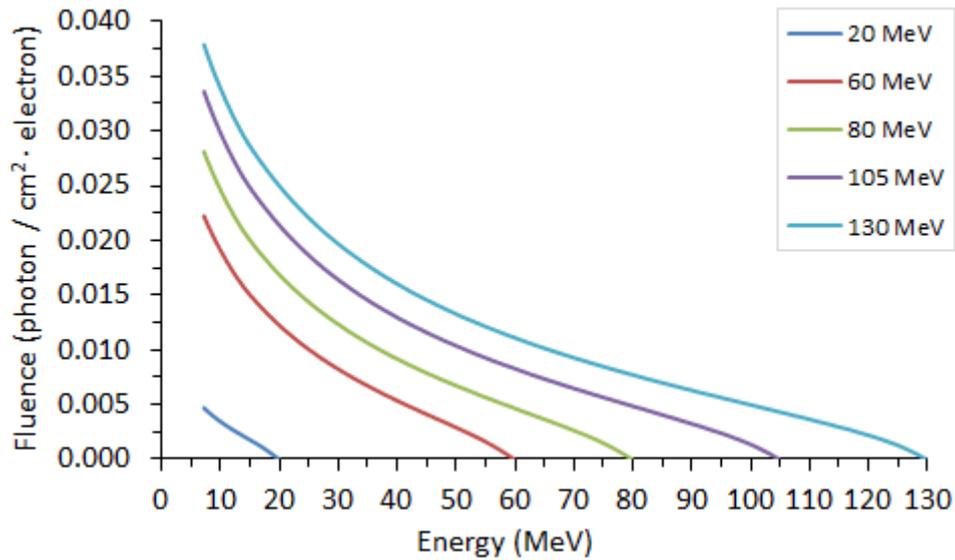

Fig.3. Bremsstrahlung fluence.

## 4. Results and discussion

### 4.1. Relative yields of reactions

Based on the results of processing the measured gamma spectra, we identified photoneutron reactions with the release of up to nine neutrons from nuclei. Table 2 shows the experimental values of the ratio of photoneutron reaction yields, and also Figure 4 shows the experimental and calculated values of the ratio of yields. Table 2 also provides literature data [4] for comparison.

Table 2. The ratio of the yields of photoneutron reactions in $^{181}$Ta nuclei. *Reaction with the formation of the isomeric state $^{178m}$Ta.

| Reactions | Energy of electrons (MeV) | | | | | | |
|---|---|---|---|---|---|---|---|
| | 20 | 40 | 60 | 67.7[4] | 80 | 105 | 130 |
| (γ,2n)/(γ,n) | 1.77(29)E-01 | 3.75(60)E-01 | 3.29(52)E-01 | 3.4(7)E-01 | 3.26(52)E-01 | 4.05(64)E-01 | 3.72(58)E-01 |
| (γ,3n)*/(γ,n) | - | 8.3(12)E-03 | 7.1(10)E-03 | 5(1)E-03 | 8.0(12)E-03 | 9.5(14)E-03 | 7.4(11)E-03 |
| (γ,3n)/(γ,n) | - | 2.06(38)E-02 | 1.72(37)E-02 | 1.8(4)E-02 | 2.25(42)E-02 | 3.04(50)E-02 | 2.93(63)E-02 |
| (γ,4n)/(γ,n) | - | 1.17(19)E-02 | 1.05(17)E-02 | 1.7(5)E-02 | 1.31(23)E-02 | 1.66(28)E-02 | 1.59(27)E-02 |
| (γ,5n)/(γ,n) | - | 3.62(54)E-04 | 3.08(45)E-03 | 5(1)E-03 | 4.32(64)E-03 | 8.5(12)E-03 | 6.17(89)E-03 |
| (γ,6n)/(γ,n) | - | - | 6.9(11)E-04 | 1.4(3)E-03 | 1.58(24)E-03 | 3.55(53)E-03 | 3.07(50)E-03 |
| (γ,7n)/(γ,n) | - | - | - | - | 3.50(56)E-04 | 1.52(23)E-03 | 1.44(21)E-03 |
| (γ,8n)/(γ,n) | - | - | - | - | - | 1.17(19)E-03 | 7.3(12)E-04 |

| (γ,9n)/(γ,n) | - | - | - | - | - | 2.82(59) E-04 | 2.70(52) E-04 |

It can be seen from Figure 4 that as the reaction threshold increases, the discrepancy between experiment and theory increases. If we do not take into account the results of the experiment with 40 MeV electrons, the discrepancies between the experimental results and calculations for the reaction ratio $^{181}$Ta(γ,2n)$^{179}$Ta / $^{181}$Ta(γ,n)$^{180}$Ta, $^{181}$Ta(γ,3n)$^{178}$Ta / $^{181}$Ta(γ,n)$^{180}$Ta, $^{181}$Ta(γ,4n)$^{177}$Ta / $^{181}$Ta(γ,n)$^{180}$Ta and $^{181}$Ta(γ,5n)$^{176}$Ta / $^{181}$Ta(γ,n)$^{180}$Ta are small and less than 50%. Starting from the reaction ratio $^{181}$Ta(γ,6n)$^{175}$Ta / $^{181}$Ta(γ,n)$^{180}$Ta to $^{181}$Ta(γ,9n)$^{172}$Ta / $^{181}$Ta(γ,n)$^{180}$Ta, the discrepancy reaches up to 4 times. And also for the reaction $^{181}$Ta(γ,3n)$^{178m}$Ta / $^{181}$Ta(γ,n)$^{180}$Ta with the formation of isomeric state of $^{178}$Ta nuclei, the discrepancy is up to 2 times. The results from the experiment with 40 MeV electrons also agree with the calculations, except for the $^{181}$Ta(γ,4n)$^{177}$Ta and $^{181}$Ta(γ,5n)$^{176}$Ta reactions, although the calculated yield ratios were calculated based on the bremsstrahlung spectrum from the tungsten converter experiments. The energy thresholds for the $^{181}$Ta(γ,4n)$^{177}$Ta and $^{181}$Ta(γ,5n)$^{176}$Ta reactions are 29.0 and 37.7 MeV, respectively, and are close to the bremsstrahlung end-point energy, and apparently theoretical models do not very well calculate the cross section in the region beyond the giant resonance at energies near the reaction threshold. In Figure 4 it can also be seen that the value of the reaction yield ratio in the experiment with electrons with an energy of 130 MeV decreases to 35% relative to the experiment with electrons with an energy of 105 MeV. Apparently this is due to the beginning of the birth of mesons. And, in the calculated values of the reaction yield ratio, such a decrease is not observed.

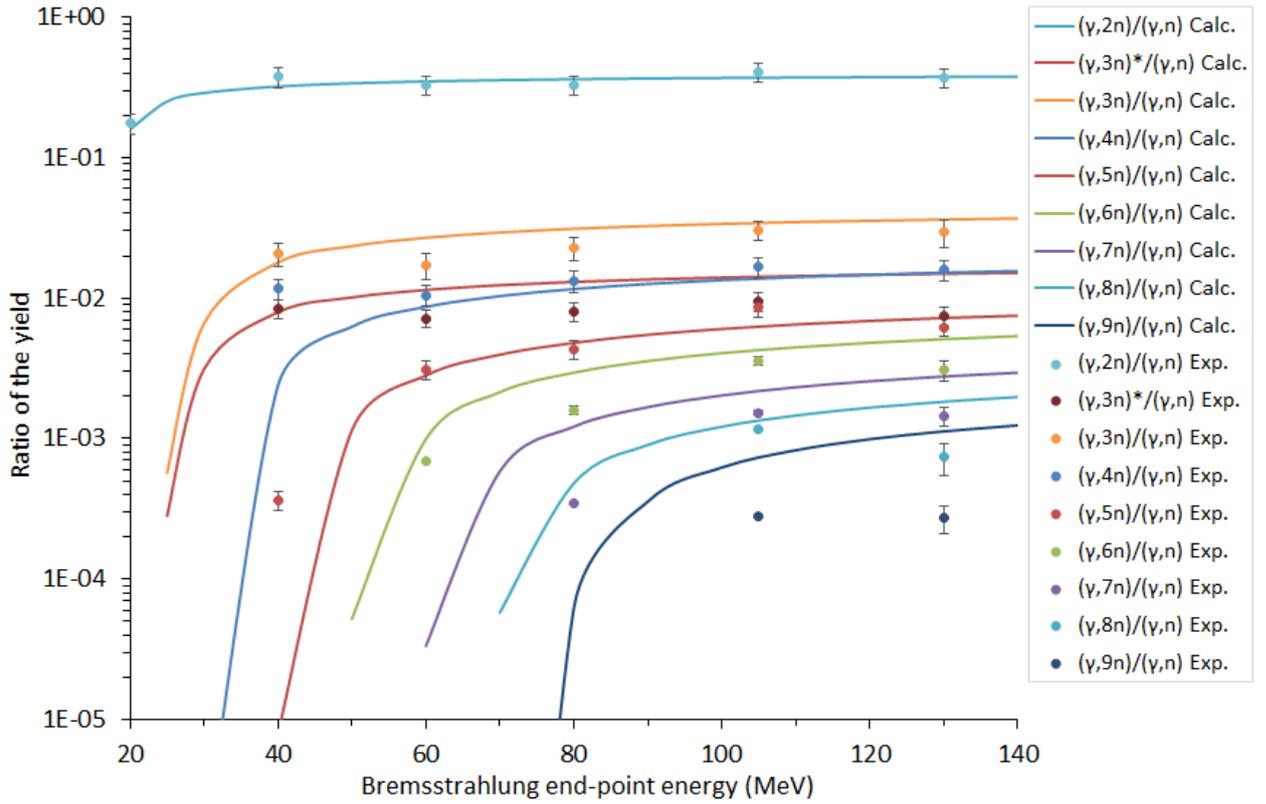

Fig.4. Experimental and calculated values of the ratio of the yields of photoneutron reactions in $^{181}$Ta nuclei. *Reaction with the formation of the isomeric state $^{178m}$Ta.

### 4.2. Flux-averaged cross-sections

Although in experiments, due to the impossibility of accurately determining the number of photons, only the relative values of the reaction yields are determined, it is still possible to determine the absolute values of the reaction yields and flux-averaged cross sections. The absolute values of the flux-averaged cross-sections make it possible to compare the obtained results with literature data. To determine the absolute values, it is necessary to select one well-studied reaction occurring within the sample being studied as a monitor, as in work [15]. As a monitor we chose the $^{181}$Ta(γ,n)$^{180}$Ta reaction, since there are several experimental data on the cross section of this reaction up to a photon energy of 35 MeV as shown in Figure 5. Having determined the difference between the experimental value of the yield of $^{181}$Ta(γ,n)$^{180}$Ta reaction and the calculated one, we used it as a correction factor to calculate the integral number of photons for all measured reactions.

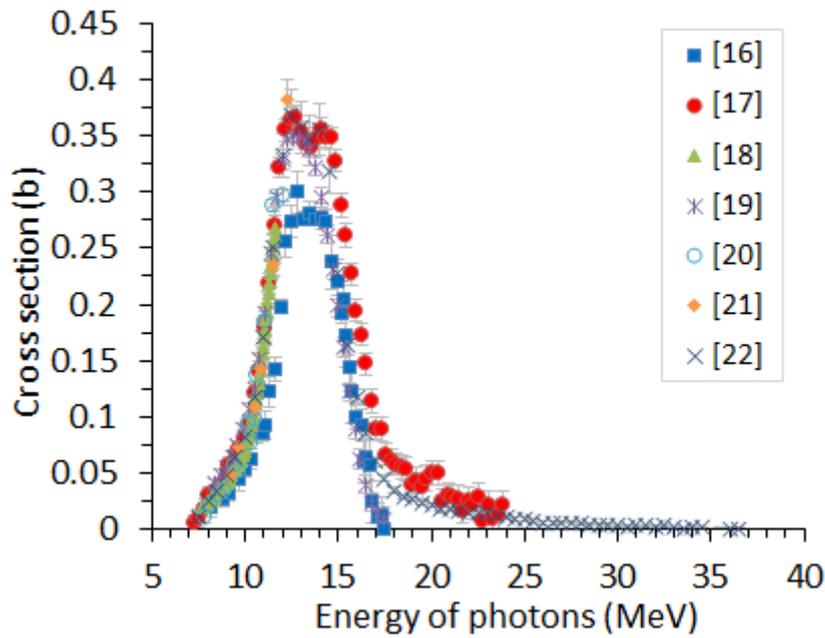

Fig.5. Cross section of the $^{181}$Ta(γ,n)$^{180}$Ta reaction.

The cross sections averaged over the bremsstrahlung flux from the threshold $E_{th}$ of the reaction under study to the end-point energy of the bremsstrahlung spectrum $E_{\gamma max}$ were calculated using formula (1) only by replacing the integral number of photons incident on the sample to the integral number of photons in the energy range $E_{th}$-$E_{\gamma max}$. Table 3 shows the flux-averaged reaction cross-section values, and Figure 6 also includes data from work [1] for comparison. From Figure 6 it can be seen that the obtained experimental data coincide with the data from work [1]. Small differences between them are related to the energy spectrum of the bremsstrahlung since the geometry of the experiment (material and thickness of the converter, etc.) was not the same, this is especially noticeable in the results of the $^{181}$Ta(γ,3n)$^{178m}$Ta reaction.

Table 3. Flux-averaged cross sections of photoneutron reactions [mb].

| Reactions | Energy of electrons (MeV) | | | | | |
|---|---|---|---|---|---|---|
| | 20 | 40 | 60 | 80 | 105 | 130 |
| $^{181}$Ta(γ,n)$^{180}$Ta | 181(21) | 104(11) | 101(11) | 92.5(99) | 87.0(92) | 73.9(76) |
| $^{181}$Ta(γ,2n)$^{179}$Ta | 420(70) | 77(12) | 65(10) | 54.4(86) | 60.5(96) | 43.8(68) |
| $^{181}$Ta(γ,3n)$^{178m}$Ta | - | 3.76(56) | 2.78(41) | 2.31(35) | 2.30(35) | 1.31(20) |
| $^{181}$Ta(γ,3n)$^{178}$Ta | - | 9.3(17) | 6.7(15) | 6.5(12) | 7.4(12) | 5.2(11) |
| $^{181}$Ta(γ,4n)$^{177}$Ta | - | 12.9(21) | 7.2(12) | 5.9(10) | 5.78(98) | 3.75(64) |
| $^{181}$Ta(γ,5n)$^{176}$Ta | - | 6.9(10) | 4.52(66) | 3.23(48) | 4.44(65) | 1.99(29) |
| $^{181}$Ta(γ,6n)$^{175}$Ta | - | - | 2.28(36) | 1.88(29) | 2.59(39) | 1.27(19) |
| $^{181}$Ta(γ,7n)$^{174}$Ta | - | - | - | 0.80(13) | 1.70(26) | 0.80(12) |
| $^{181}$Ta(γ,8n)$^{173}$Ta | - | - | - | - | 1.88(16) | 0.51(16) |
| $^{181}$Ta(γ,9n)$^{172}$Ta | - | - | - | - | 0.77(16) | 0.26(5) |

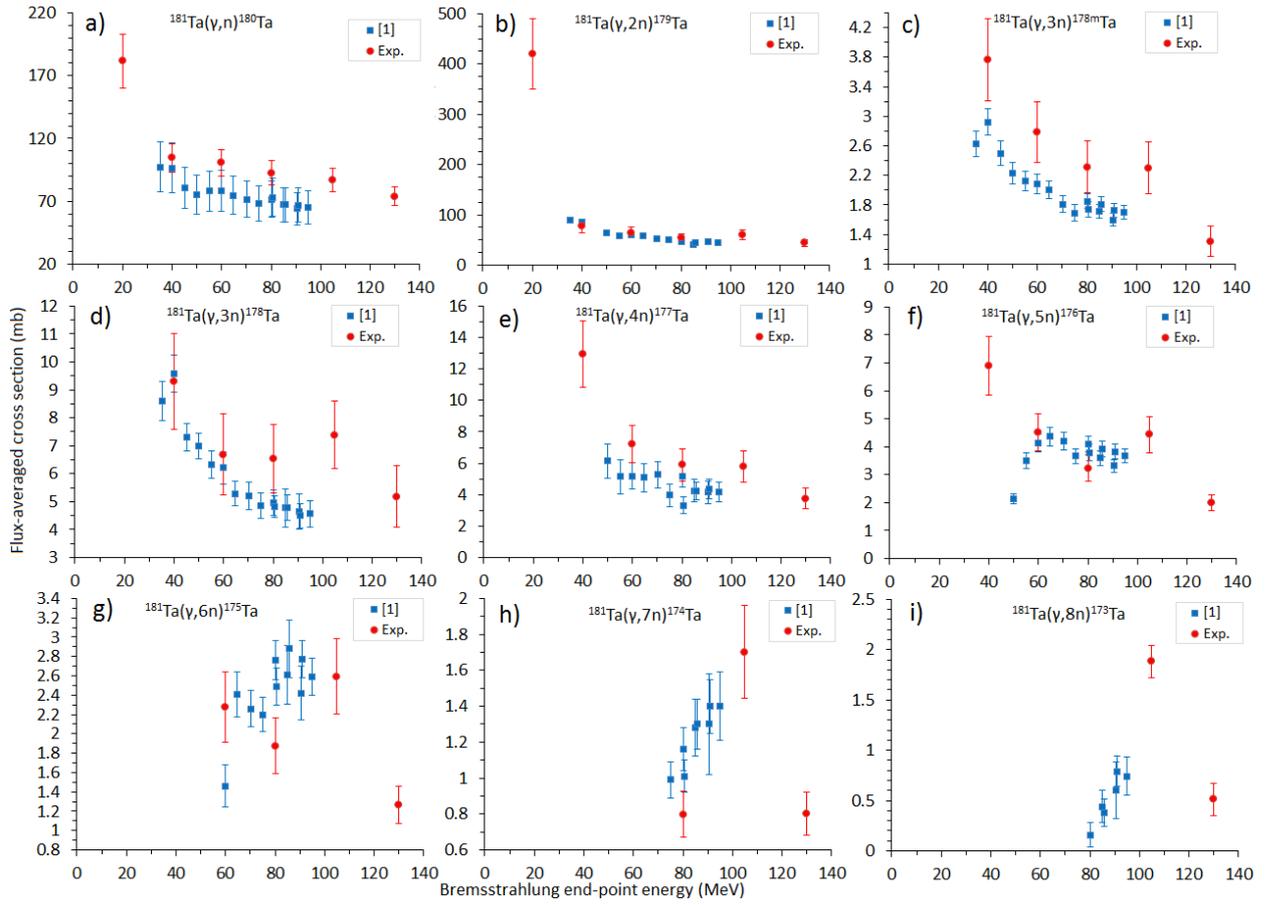

Fig. 6. Flux-averaged cross sections of photoneutron reactions in $^{181}$Ta nuclei.

### 4.3. Uncertainty of the results

The measurement uncertainty of experimental values of reaction yield ratios was determined as the squared sum of the standard deviation of the arithmetic mean of the relative yield values and systematical errors. The standard deviation of the arithmetic mean of relative yield values also includes statistical errors. The statistical error of the observed γ-rays is mainly due to the statistical calculation of the total absorption peak of the corresponding γ-line, which varies between 1–10%. Systematical errors include errors in detector efficiency, half-lives of residual nuclei, and the intensity of γ-rays emitted during the decay of residual nuclei. The detection efficiency of γ-radiation of the HPGe detector is up to 5%, and this value is associated with the error in the activity of standard γ-radiation sources. The half-life T1/2 of the residual nuclei and the intensity of the analyzed γ-rays have an error of 1 to 20%, as indicated in Table 1 according to data from [10]. In this work, we did not take into account the errors in the accelerated electron beam current, since we determined the value of the ratio of the yields of photoneutron reactions. The uncertainty of the values of flux-averaged cross-sections is equated to the uncertainty of the values of relative yields.

## 5. Conclusion

The bremsstrahlung flux was formed by irradiating a tungsten converter with an electron beam from the LINAC-200 accelerator. The experiments were carried out in electron beams with energies of 20, 40, 60, 80, 105 and 130 MeV, and in each experiment the tantalum sample was irradiated with a bremsstrahlung flux. As a result of processing the obtained data, multinucleon photonuclear reactions with the emission of up to 9 neutrons in $^{181}$Ta nuclei were identified. For each identified photonuclear reaction, relative reaction yields and flux-averaged cross-sections were determined. Calculation studies of the yields of photonuclear reactions were performed using the Geant4 and TALYS-2.0 codes. The bremsstrahlung fluxes for tantalum samples were calculated with the Geant4 code, and the photonuclear reaction cross-sections were calculated with the TALYS-2.0 code.

The obtained experimental results were compared with available literature data and calculated results. The experimental values of the relative reaction yield and flux-averaged cross-section coincide with the literature data, taking into account the different geometry of the experiments. The differences between the experimental and calculated values of the relative yields are less than 50% in reactions with up to 5 neutrons emitted from the nucleus, but then, with an increase in the number of neutrons emitted from the nucleus, the difference between the experiment and the calculation also increases. These comparisons show that the used theoretical models are not able to correctly describe photonuclear reactions with the emission of more than 5-6 neutrons from the nucleus.

**Acknowledgments**

The authors express their gratitude to the team of the LINAC-200 electron accelerator of the Joint Institute for Nuclear Research for cooperation in the implementation of the experiments.

**References**

1. O.S. Deiev, I.S. Timchenko, S.N. Olejnik, V.A. Kushnir, V.V. Mytrochenko, and S.A. Perezhogin. Cross sections of photoneutron reactions on 181Ta at $E_{\gamma\text{max}}$ up to 95 MeV. Phys. Rev. C 106, 024617 (2022) https://doi.org/10.1103/PhysRevC.106.024617
2. M.B. Chadwick, P. Obloinský, P.E. Hodgson, and G. Reffo, Pauli-blocking in the quasideuteron model of photoabsorption, Phys. Rev. C 44, 814 (1991). https://doi.org/10.1103/PhysRevC.44.814
3. B.S. Ishkhanov and V.N. Orlin, Combined model of photonucleon reactions, Phys. At. Nucl. 74, 19 (2011). https://doi.org/10.1134/S1063778811010054

4. B.S. Ishkhanov, V.N. Orlin, and S. Yu. Troschiev, Photodisintegration of tantalum, Phys. At. Nucl. 75, 253 (2012). https://doi.org/10.1134/S1063778812020093
5. B.S. Ishkhanov, I.M. Kapitonov, A.A. Kuznetsov, V.N. Orlin, and H.D. Yoon, Photodisintegration of molybdenum isotopes, Moscow Univ. Phys. Bull. Ser. 3 Fiz. Astron. 1, 35 (2014), in Russian.
6. T.E. Rodrigues, J.D.T. Arruda-Neto, A. Deppman, V.P. Likhachev, J. Mesa, C. Garcia, K. Shtejer, G. Silva, S.B. Duarte, and O.A.P. Tavares, Photonuclear reactions at intermediate energies investigated via the Monte Carlo multicollisional intranuclear cascade model, Phys. Rev. C 69, 064611 (2004). https://doi.org/10.1103/PhysRevC.69.064611
7. A. Leprêtre, H. Beil, R. Bergère, P. Carlos, J. Fagot, A. De Miniac, and A. Veyssière, Measurements of the total photonuclear cross sections from 30 MeV to 140 MeV for Sn, Ce, Ta, Pb and U nuclei, Nucl. Phys. A 367, 237 (1981). https://doi.org/10.1016/0375-9474(81)90516-9
8. M.A. Nozdrin, V.V. Kobets, R.V. Timonin, et al. Design of the New Control System for Linac-200, Physics of Particles and Nuclei Letters 17, 600-603 (2020). https://link.springer.com/article/10.1134/S1547477120040342
9. J. Frána, Program DEIMOS32 for gamma-ray spectra evaluation, J. Radioanal. Nucl. Chem. 257 (3) (2003) 583–587, http://dx.doi.org/10.1023/A:1025448800782
10. S.Y.F. Chu, L.P. Ekstrom, R.B. Firestone, The Lund/LBNL Nuclear Data Search, Version 2.0, February 1999, WWW Table of Radioactive Isotopes, http://nucleardata.nuclear.lu.se/toi/
11. A.J. Koning, S. Hilaire, S. Goriely, TALYS: modeling of nuclear reactions, Eur. Phys. J. A 59, 131 (2023), https://doi.org/10.1140/epja/s10050-023-01034-3
12. J. Allison, et al., Recent developments in Geant4, Nucl. Instrum. Methods Phys. Res. A 835 (2016) 186–225, http://dx.doi.org/10.1016/j.nima.2016.06.125
13. S.M. Seltzer and M.J. Berger. Bremsstrahlung spectra from electron interactions with screened atomic nuclei and orbital electrons. Nucl. Instr. and Meth. in Phys. Research B, 12, 95-134 (1985), https://doi.org/10.1016/0168-583X(85)90707-4
14. S.M. Seltzer and M.J. Berger. Bremsstrahlung energy spectra from electrons with kinetic energy 1 kev - 100 gev incident on screened nuclei and orbital electrons of neutral atoms with z = 1–100. Atomic Data and Nuclear Data Tables, 35, 345-418, (1986), https://doi.org/10.1016/0092-640X(86)90014-8
15. Remizov, P.D., Zheltonozhskaya, M.V., Chernyaev, A.P. et al. Measurements of the flux-weighted yields for (γ, αXn) reactions on molybdenum and niobium. Eur. Phys. J. A 59, 141 (2023). https://doi.org/10.1140/epja/s10050-023-01055-y

16. R. L. Bramblett, J. T. Caldwell, G. F. Auchampaugh, and S. C. Fultz. Photoneutron Cross Sections of Ta181 and Ho165. Phys.Rev. 129 (1963) 2723-2729. https://doi.org/10.1103/PhysRev.129.2723
17. R. Bergère, H. Beil, A. Veyssière, Photoneutron cross sections of La, Tb, Ho and Ta, Nuclear Physics A, 121, 2, 463-480 (1968) https://doi.org/10.1016/0375-9474(68)90433-8
18. S.N. Belyaev, A.B. Kozin, A.A. Nechkin, V.A. Semenov, S.F. Semenko. Photoabsorption cross sections in the isotopes of Pr, Bi, and Ta in the energy region $E_{gamma}$ < 12 MeV. Soviet Journal of Nuclear Physics, volume 42, page 662, 1985.
19. V.V. Varlamov, N.N. Peskov, D.S. Rudenko, M.E. Stepanov. Photoneutron reaction cross sections in experiments with beams of quasimonoenergetic annihilation photons. Jour. Vop. At. Nauki i Tekhn., Ser. Yaderno-Reak. Konstanty, Issue.1-2, p.48, 2003.
20. H. Utsunomiya, H. Akimune, S. Goko, M. Ohta, H. Ueda, T. Yamagata, K. Yamasaki, H. Ohgaki, H. Toyokawa, Y.-W. Lui, T. Hayakawa, T. Shizuma, E. Khan, and S. Goriely. Cross section measurements of the 181Ta($\gamma,n$)180Ta reaction near neutron threshold and the p-process nucleosynthesis. Phys. Rev. C 67, 015807 (2003). https://doi.org/10.1103/PhysRevC.67.015807
21. Goko S, Utsunomiya H, Goriely S, Makinaga A, Kaihori T, Hohara S, Akimune H, Yamagata T, Lui YW, Toyokawa H, Koning AJ, Hilaire S. Partial photoneutron cross sections for the isomeric state 180mTa. Phys. Rev. Lett. 96, 192501 (2006). https://doi.org/10.1103/PhysRevLett.96.192501
22. Varlamov, V.V., Ishkhanov, B.S., Orlin, V.N. et al. New data on (γ,n), (γ,2n), and (γ,3n) partial photoneutron reactions. Phys. Atom. Nuclei 76, 1403–1414 (2013). https://doi.org/10.1134/S1063778813110148